\documentclass[fleqn,10pt]{wlscirep}
\usepackage[utf8]{inputenc}
\usepackage[T1]{fontenc}
\usepackage{mathtools}
\usepackage{braket}
\usepackage{graphicx}
\usepackage{amsmath}
\usepackage{url}

\title{Quantum Teleportation with One Classical Bit}

\author{Abhishek Parakh}
\affil{University of Nebraska, Omaha, NE 68116}


\begin{abstract}
Quantum teleportation allows one to transmit an arbitrary qubit from point A to point B using a pair of (pre-shared) entangled qubits and classical bits of information. The conventional protocol for teleportation uses two bits of classical information and assumes that the sender has access to only one copy of the arbitrary qubit to the sent. Here, we ask whether we can do better than two bits of classical information if the sender has access to multiple copies of the qubit to be teleported. We place no restrictions on the qubit states. Consequently, we propose a modified quantum teleportation protocol that allows Alice to reset the state of the entangled pair to its initial state using only local operations. As a result, the proposed teleportation protocol requires the transmission of only one classical bit with a probability greater than one-half. This has implications for efficient quantum communications and security of quantum cryptographic protocols based on quantum entanglement.
\end{abstract}
\begin{document}

\flushbottom
\maketitle
%
%
\thispagestyle{empty}


\section*{Introduction}

Quantum teleportation \cite{orgTeleportation1993, Gauthier2009} is a strikingly curious quantum phenomenon with myriads of applications ranging from secure quantum communication to distributed quantum computing \cite{quantumRepeater1998, quantumComputation1999,quantumComputation2-2001,quantumInternet2008, universalQCTeleport1999}. First presented by Bennett et al. \cite{orgTeleportation1993}, quantum teleportation allows one to recreate an arbitrary qubit from just two bits of classical information. The precondition is that the two communicating parties share an entangled pair of qubits. Entanglement, essentially, transmits the raw values of amplitudes present in the arbitrary qubit and the classical bits provide a final ``correction'' to these values. Quantum teleportation has proven to be an invaluable tool in quantum information science \cite{quantumTeleQIS2012, quantumTeleportQIS2005, wildeBook_2017}. Quantum teleportation enables the development of quantum repeaters \cite{quantumRepeater1998, quantumRepeater2020}, a pivotal technology for the establishment of quantum communication networks \cite{QNIA2018}, and in general the framework of quantum Internet. Quantum teleportation has been implemented in lab setting using a number of different resources \cite{teleportImplement2020, teleImplement2021}. Lastly, a number of cryptographic protocols use shared entangled qubits and operations equivalent to teleportation to produce random numbers and encryption keys \cite{qkdReview2017}.

It is known that, given a pre-shared Bell pair, the teleportation of an unknown qubit, requires the transmission of two classical bits \cite{orgTeleportation1993}. More generally, the teleportation of an $N$-state requires 2$log_2 N$ classical bits \cite{teleportCost1996}. This, however, assumes that the sender has access to only one copy of the arbitrary unknown qubit that she is teleporting. Kak investigated the minimum number of classical bits required for teleportation \cite{Kak2003TeleportationPR}. In turn, he proposed three variations on Bennett's protocol that required fewer than two qubits. Two of these variations, however, use a non-traditional setup where the entangled qubits are not pre-shared between the communicating parties but transmitted along with the classical bits of information. This change in the setup can be looked upon as an encryption system where the classical bit of information acts as the encryption key for the qubit and multipath routing can be used for their transmission. From the viewpoint of teleportation, nevertheless, if the qubits are to be transmitted along with classical bits, then one could transmit the arbitrary qubit directly at that time without the need to resorting to teleportation. However, it does point out to the possibility of some ``non-standard'' setting of quantum teleportation having lower classical cost of communication. One such ``non-standard'' setting was investigated by Pati \cite{remoteStatePreparation2000} where the teleported qubit arose from a specific portion of the Bloch sphere leading to a one-bit classical communication cost. This was further investigated by Lo as remote state preparation \cite{teleportTwoStage2000}.

In this paper, we too consider a somewhat non-standard setting but in terms of resources. Like the traditional teleportation protocol, we assume that the communicating parties pre-share entangled qubits. However, we consider the case when the sender has more than one copy of an (unknown) arbitrary qubit $\ket{\phi}=a\ket{0}+b\ket{1}$, $|a|^2+|b|^2=1$. We do not place any other restrictions on the qubits. Such a possibility has not been investigated before. Our result shows that under such as setting, we can do better than two classical bits of communication. A sender may have multiple copies of the qubit to be teleported in many practical situations where the unknown qubit is either the result of a periodic or on-demand natural process or output of a quantum algorithm. One may need to send this (unmeasured) output to the receiver for further processing. We show that the teleportation of such a qubit requires the transmission of only one classical bit with a probability greater than one-half.  In order to do so, we propose a \textit{reset} procedure that allows for the recreation of the original shared Bell state using only local operations at the sender's end and no consumption of resources at receiver's end. In the proposed reset procedure, as with conventional teleportation protocol, we begin with maximally entangled qubits. At the first stage of the protocol, if this maximally entangled qubit enters into a desired state we move on to the second stage of the protocol. If, on the other hand, the entangled part ends up in an undesired state we reset it back to the original state using only local operations at the sender's end. The cost of doing so is one copy of the arbitrary qubit, $\ket{\phi}$.

Furthermore, if the amplitudes $a$ and $b$ are known and/or the qubit, $\ket{\phi}$, generated on demand, then the \textit{reset} operation may be attempted many times over.

The result presented in this paper, we believe, represents a fundamental advance in the classical burden of teleportation, since it was proposed by Bennett et al.\cite{orgTeleportation1993}, because we do not restrict the states of the qubit and allow the parties to pre-share entangled pairs as in the traditional setting. The presented result also has widespread consequences for quantum cryptographic protocols that use entangled pairs and assume a symmetric power between communicating parties.

\section{Result}

It is useful to think of the conventional teleportation protocol as a two stage system \cite{teleportTwoStage2000}. We start with a shared Bell pair $\frac{\ket{00}+\ket{11}}{\sqrt{2}}$ between sender, Alice, and receiver, Bob. The first stage of the teleportation protocol introduces the amplitudes $a$ and $b$ into the entangled pair. However, the locations of the amplitudes maybe flipped and is corrected with the application of Pauli X gate by both the sender and the receiver. This requires the transmission of 1-bit of classical information to Bob. In the second stage of the protocol, Alice terminates the entanglement by measuring her qubit of the Bell pair in the Hadamard basis. This introduces another undesirable error into the state of the qubit that Bob holds corresponding to the Pauli Z gate. This is corrected by the transmission of another classical bit of information to Bob.

To be more precise, in Bennett's protocol \cite{orgTeleportation1993} at the end of the first stage the two parties share the state $a\ket{00}+b\ket{11}$ or $b\ket{00}+a\ket{11}$. The first of these states is desirable and the second one is converted to the first one with the application of a bi-local unitary transformation $\ket{0}\rightarrow\ket{1}$. This requires one bit to be sent to Bob. In the second stage of the protocol, Bob ends up with $a\ket{0}+b\ket{1}$ or $a\ket{0}-b\ket{1}$. The correction of the last state requires the other bit of information to be sent to Bob. In the end Bob is left with $a\ket{0}+b\ket{1}$, which corresponds to the unknown state $\ket{\phi}$ that Alice wanted to teleport.

In this paper, we ask the simple question: if Alice ends up in the undesired state at the end of the first stage, i.e. $b\ket{00}+a\ket{11}$, can she using only local transformations, reset the state of the shared entanglement to the original state $\frac{\ket{00}+\ket{11}}{\sqrt{2}}$? If so, she can reattempt the teleportation protocol.

\subsection{Resetting the Entangled State} 

In order to reset the entangled state, Alice introduces an ancillary qubit $\ket{\theta}=c\ket{0}+d\ket{1}$ into the system. It will be come apparent through the following discussion that $\ket{\theta}=\ket{\phi}$. The resulting system state is given by $\ket{\psi}$,

\begin{equation}
\begin{split}
    \ket{\psi}&=\ket{\theta}(a\ket{1_A1_B}+b\ket{0_A0_B})\\
    &=(c\ket{0_l}+d\ket{1_l})(a\ket{1_A1_B}+b\ket{0_A0_B})\\
    &=ca\ket{0_l 1_A 1_B}+cb\ket{0_l 0_A 0_B}+da\ket{1_l 1_A 1_B}+db\ket{1_l 0_A 0_B}
\end{split}
\end{equation}

Qubits with subscripts $l$ and $A$ are with Alice and the qubit denoted by subscript $B$ is with Bob. Now, Alice applies CNOT with the ancillary qubit (the first qubit) as the control qubit and second qubit as the target qubit. The state then becomes,

\begin{equation}
\begin{split}
    \ket{\psi}&=ca\ket{0_l 1_A 1_B}+cb\ket{0_l 0_A 0_B}+da\ket{1_l 0_A 1_B}+db\ket{1_l 1_A 0_B}\\
    &=\ket{1_A}(ca\ket{0_l 1_B}+db\ket{1_l 0_B}) + \ket{0_A}(cb\ket{0_l 0_B}+da\ket{1_l 1_B})
    \label{eq:psi_4}
\end{split}
\end{equation}

In Equation \ref{eq:psi_4} above, we have removed the second qubit as the common qubit for simplification of the expression. We see that $\ket{0_A}(cb\ket{0_l 0_B}+da\ket{1_l 1_B})$ is closest to the desired state $\frac{\ket{00}+\ket{11}}{\sqrt{2}}$. If we set $c=a$ and $d=b$ then we get,
\begin{equation}
    \ket{\psi}=\ket{1_A}(aa\ket{0_l 1_B}+bb\ket{1_l 0_B}) + \ket{0_A}(ab\ket{0_l 0_B}+ba\ket{1_l 1_B})
\end{equation}

Alice now measures the second qubit. 

She will see $\ket{0}$ with probability $2|ab|^2$ and the shared entangled qubits end up in state $\frac{\ket{00}+\ket{11}}{\sqrt{2}}$. \textit{This constitutes a successful reset by Alice}. Note that, setting $c=a$ and $d=b$ means that Alice is simply using a copy of the original qubit $\ket{\phi}$ for the reset operation and does not need to create any new special ancillary qubits. Once a reset to the original Bell state is achieved, Alice reinitiates the teleportation protocol using a fresh copy of $\ket{\phi}$.

On the other hand, Alice's measurement results in $\ket{1}$ with probability $|a^2|^2+|b^2|^2$. This takes the system to state $\frac{a^2\ket{01}+b^2\ket{10}}{\sqrt{|a^2|^2+|b^2|^2}}$. Alice can apply the $X$ gate to her qubit and reattempt the reset operation with $\frac{a^2}{\sqrt{|a^2|^2+|b^2|^2}}$ and $\frac{b^2}{\sqrt{|a^2|^2+|b^2|^2}}$ as the new $a$ and $b$ values, respectively.

\begin{figure}[ht]
    \centering
    \includegraphics[width=0.7\textwidth]{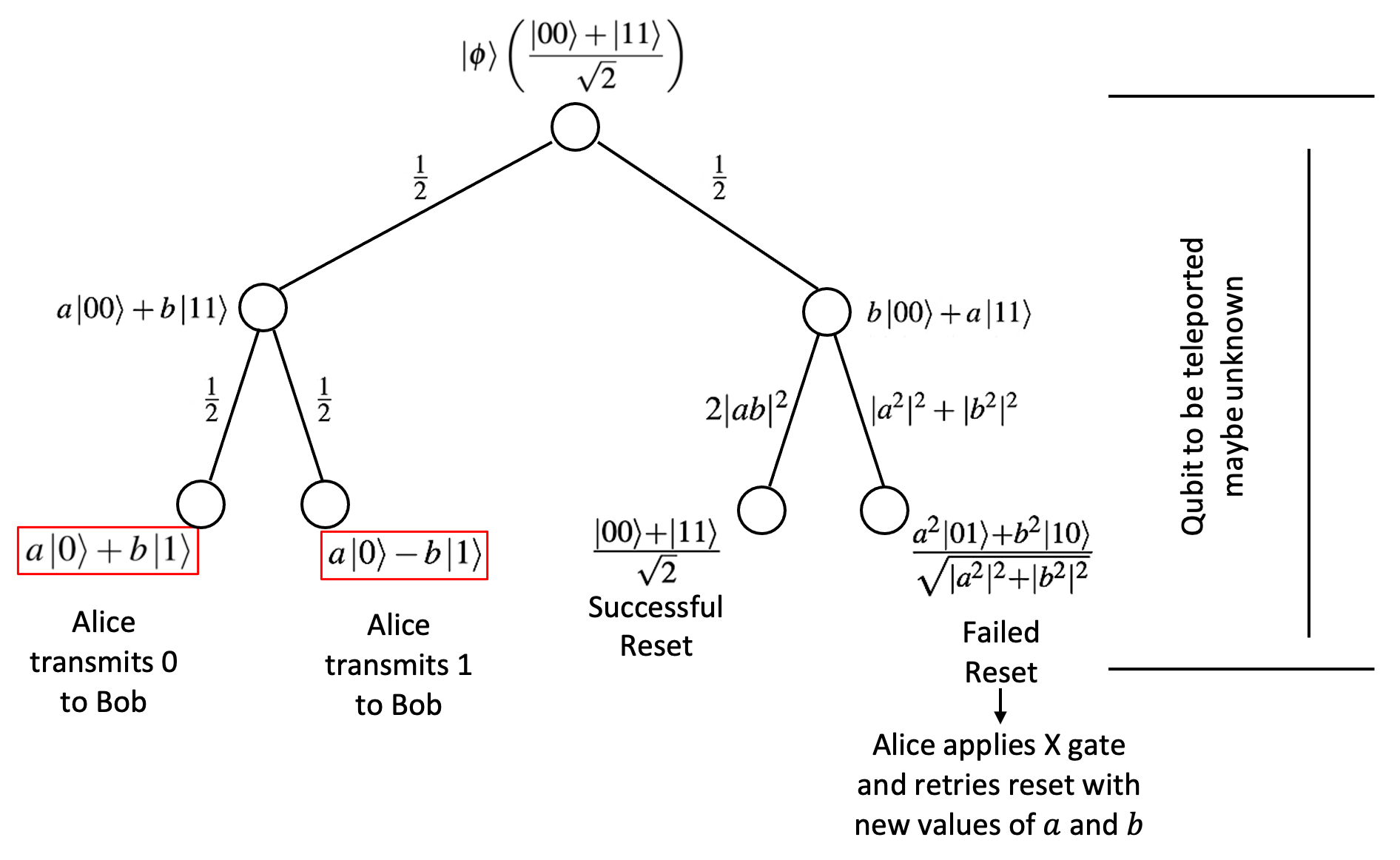}
    \caption{Decision tree at Alice's end for teleportation with the proposed reset operation. Probabilities are shown along the arms between the nodes. States in the red boxes reside with Bob.}
    \label{fig:teleportTree}
\end{figure}

At this point, Alice has consumed two copies of $\ket{\phi}$. One during stage one of the teleportation protocol and second as the ancillary qubit introduced into the system during the reset operation. Figure \ref{fig:teleportTree} shows the branches that a teleportation attempt with the proposed reset procedure can follow.

\section{Discussion}

The probability of success of the reset attempts vary with the values of amplitudes $a$ and $b$, in the state $\ket{\phi}$, and is given by $2|ab|^2$ for the first reset attempt. Since, $|ab|^2=|a|^2|b|^2 = |a|^2(1-|a|^2) = |a|^2-|a|^4$ and $0\leq |a|^2\leq 1$, we get $|ab|^2\leq \frac{1}{4}$ and $2|ab|^2\leq \frac{1}{2}$.

For example, when $a=\frac{i}{\sqrt{2}}$ and $b=\frac{1}{2}+\frac{i}{2}$, the probability that the first reset attempt will succeed is $\frac{1}{2}$. After a successful reset, Alice makes a second attempt to teleport the qubit. The probability that entangled qubits will collapse into the desirable state ($a\ket{00}+b\ket{11}$) is again $\frac{1}{2}$. Therefore, with probability $\frac{1}{2}+\frac{1}{4}$ Alice would transmit just one bit for teleportation but must have at least three copies of $\ket{\phi}$ at her disposal. 

Remarkably, the qubit $\ket{\phi}$ remains unknown and arbitrary.

Consequently, in general, considering only the first reset attempt and an unknown qubit $\ket{\phi}$ the probability of ending up in the desired state ($a\ket{00}+b\ket{11}$), that is a successful teleportation with only one-classical bit, is $\frac{1}{2}+\frac{1}{2}\cdot2|ab|^2=\frac{1}{2}+|ab|^2$. Note, this probability is always set to $\frac{1}{2}$ in Bennett et al.'s quantum teleportation \cite{orgTeleportation1993} protocol which always requires two bits of information to be transmitted.

Alice, however, does require the values of $a$ and $b$ for the reset attempts beyond the first one; in the case of a reset failure. This latter case is useful when the precision required for classical representation of the $a$ and $b$ values exceeds the available bandwidth or resources available on the classical channel. In other words, if the values of $a$ and $b$ are known, Alice can pursue the reset attempts until it succeeds. Note that once a reset attempt succeeds, the probability that we will end up in the desired entangled state is again $\frac{1}{2}$. Therefore, the expected number of attempts, given a successful reset, is 2.

The reset procedure presented above, when successful, reduces the teleportation protocol to one stage protocol under the standard setting of pre-shared Bell states. Furthermore, Alice knows at each stage whether the reset has succeeded or not. In the worst case, if the values of $a$ and $b$ are not known, and the reset fails, she can abandon that particular pair of entangled qubits and use another. The computational burden of teleportation with reset can then be stated as,

\begin{equation}\label{eq:entropy}
    H(T)=\left[(\frac{1}{2}+|ab|^2)(1)+ (\frac{1}{2}-|ab|^2) (2)\right] \mbox{bits}    
\end{equation}

It is easy to see that $1.25\leq H(T)\leq 1.5$ depending on the values of $a$ and $b$. For comparison, one of the protocols by Kak \cite{Kak2003TeleportationPR} reduces the computational burden to 1.5 bits when the Bell states are pre-shared. The cost of reducing the computational burden to 1.25 bits, therefore, is the expenditure of extra copies of the unknown qubit $\ket{\phi}$. In other words, the probability $\frac{1}{2}+|ab|^2$ only represents the average case. Given that the proposed protocol is probabilistic in nature and if Alice has several qubits to teleport, she may end up with a situation where all her reset attempts may succeed resulting in a 1 bit per qubit for teleportation even for a completely arbitrary unknown qubit! 

One may argue that even in Bennett's original teleportation protocol \cite{orgTeleportation1993}, Alice and Bob may agree on a scheme such as the follows: if stage one of the protocol succeeds in creating $a\ket{00}+b\ket{11}$ then they will proceed with stage two and if stage one results in $a\ket{11}+b\ket{00}$, they will abandon the entangled pair. Such a modification would also be probabilistic and result in 1 bit per qubit for teleportation. However, we point out that the probability of 1 bit per qubit in such a modified Bennett's protocol would be stuck at $\frac{1}{2}$. Whereas, our reset procedure allows this probability to be $\frac{1}{2}+|ab|^2$ which is greater than $\frac{1}{2}$ for all practical purposes.

The probability of success for the reset operation at every attempt changes with the amplitudes $a$ and $b$ and is given by a recursive relationship. If $R_0, R_1, R_2, \ldots$ represents the reset successive reset attempts then the probability of success of each reset is given by,

\begin{equation}
\begin{split}
& P(R_0)=2|ab|^2\\
&P(R_1)=2|a_1b_1|^2 \mbox{ where } a_1=\frac{a^2}{\sqrt{|a^2|^2+|b^2|^2}} \mbox{ and } b_1=\frac{b^2}{\sqrt{|a^2|^2+|b^2|^2}}\\
&P(R_2)=2|a_2b_2|^2 \mbox{ where } a_2=\frac{a_1^2}{\sqrt{|a_1^2|^2+|b_1^2|^2}} \mbox{ and } b_2=\frac{b_1^2}{\sqrt{|a_1^2|^2+|b_1^2|^2}}\\
&\mbox{and so on.}
\end{split}
\end{equation}

Therefore, the probability that Alice would \textit{need} three reset attempts is given by $(1-P(R_0))(1-P(R_1))P(R_2)$. For a state such as $\frac{\ket{0}+\ket{1}}{\sqrt{2}}$ this would be $\frac{1}{8}$.

An important implication of Equation \ref{eq:entropy} is that when $|ab|^2> 0$ there is a non-zero probability of a successful reset. The equality, $|ab|^2=0$, holds only when one of the amplitudes $a$ or $b$ = 0, a corner case. As a result, the above protocol breaks the symmetry between the four cases of the conventional teleportation protocol \cite{orgTeleportation1993}. More precisely, in the conventional teleportation protocol, there is an equal probability for Bob to end up in any of the four states: $\ket{\phi_0}=a\ket{0}+b\ket{1}$, $\ket{\phi_1}=a\ket{0}-b\ket{1}$, $\ket{\phi_2}=b\ket{0}+a\ket{1}$, and $\ket{\phi_3}=b\ket{0}-a\ket{1}$. Consequently, two classical bits were needed for teleportation. However, with the proposed reset operation the relationship between the probabilities of the states that Bob sees is given by, 
\begin{equation}
P(\ket{\phi_0})+P(\ket{\phi_1})>P(\ket{\phi_2})+P(\ket{\phi_3})    
\end{equation}

For a moment only consider the first stage of the teleportation protocol and the reset operation. Assume that Alice has not transmitted any information to Bob. We see that, the reset operation also implies that when $a, b\neq 0$, Alice can unilaterally force the shared entanglement, $\frac{\ket{00}+\ket{11}}{\sqrt{2}}$ to be converted to $a\ket{00}+b\ket{11}$ with a probability higher than $\frac{1}{2}$ and no communication with Bob. If the qubit $\ket{\phi}$ is deliberately constructed by Alice and the information Alice wants to transmit is encoded in the values of $a$ and $b$ rather than the phase difference between them, then Alice has successfully induced a bias in Bob's qubit. 

The ability of one party to, unilaterally, induce a bias in a shared entangled pair gives a party wanting to cheat, in a cryptographic protocol, an advantage. For example, several quantum key agreement protocols have been proposed in literature \cite{qka2004, qka2017, Yin2013ThreePartyQK, qka2019, qka2011, qka2013}. These protocols are designed such that both the communicating parties make equal contribution to the final agreed key. This is in contrast to traditional quantum key distribution protocols such as BB84 where one party determines the entire key and securely distributes it to the other party. If the quantum key agreement protocol is based on entanglement then one of the cheating parties can induce a bias in the final key stream by introducing an appropriate $\ket{\phi}$ in the system. The simplest example of a weak quantum key agreement protocol is the E91 protocol where both parties make random measurements on maximally entangled pairs of qubits. If these maximally entangled pairs are distributed by a central authority then Alice (cheating party) can introduce $\ket{\phi}=a\ket{0}+b\ket{1}$ such that $|a|^2>>|b|^2$ resulting in a biased measurement result, at Bob's end, of her own choosing. Therefore, the asymmetry caused by the reset operation gives Alice more power than Bob and is devastating for cryptographic protocols that use entanglement and rely on the powers of the communicating parties being symmetric for security. 

In this paper, we have only considered the ideal case where the shared entangled pair is maximally entangled. Investigation into how the reset procedure would proceed in the case of non-maximally entangled states, mixed states, non-ideal measurements and channel errors are left as future work. Entanglement distillation \cite{distill1996} could be used to mitigate the effects of these non-ideal cases.

\bibliography{main}



\end{document}